\documentclass{webofc}

\usepackage[varg]{txfonts}   
\usepackage{hyperref}
\usepackage{url}
\RequirePackage{orcidlink}

\hypersetup{colorlinks=true,citecolor=blue,urlcolor=blue,linkcolor=blue}
\begin{document}
\title{Usage of GPUs for ALICE Run 3 Offline Reconstruction on the GRID}
%
%

\author{\firstname{David} \lastname{Rohr}\inst{1}\orcidlink{0000-0003-4101-0160}\fnsep\thanks{\email{drohr@cern.ch}} for the ALICE Collaboration
}

\institute{CERN, Esplanade des Particules 1, 1217 Meyrin, Switzerland
          }

\abstract{ALICE, the dedicated heavy ion experiment at the LHC, records Pb--Pb collisions at an interaction rate of up to 50 kHz.
It was the first LHC experiment to leverage GPUs for online data processing during LHC Runs 1 and 2, and its Run 3 online data processing is now fully based on GPUs with more than 90\% of the compute load offloaded to the accelerator.
In order to use its online processing server farm also for offline processing in an efficient way while the LHC is not operating, ALICE has been running the offline TPC tracking on GPUs since 2023.
Since then, ALICE has been conducting an ongoing effort to offload more offline reconstruction steps to GPUs, and to use the GPUs at other GRID sites beyond the ALICE online computing farm for offline reconstruction.
In particular, porting the track model decoding and ITS tracking to GPU has improved the throughput by 29\%.
In 2026, ALICE ran the first GPU reconstruction jobs on the NVIDIA GPUs of the NERSC Perlmutter cluster, marking the first use of GPUs for LHC offline reconstruction on the GRID.
Users can submit GPU reconstruction jobs by specifying only the required GPU count and type, while the underlying infrastructure handles all remaining configurations transparently.
}
\maketitle
\section{Introduction}

ALICE~\cite{bib:alice} (A Large Ion Collider Experiment) is the dedicated heavy-ion experiment at the LHC (Large Hadron Collider) at CERN.
For Run 3, ALICE upgraded its detectors and computing scheme to cope with much higher rates~\cite{bib:ls2upgrade, bib:o2tdr}.
In order to access low signal over background effects, ALICE stores all recorded Pb--Pb collisions and operates without a trigger but with continuous readout.
Due to ALICE's slow drift-detectors like the TPC (Time Projection Chamber), collisions from multiple bunch crossings overlap in the detector.
It is thus impossible to process individual collisions, but the assignment of clusters to a collision can only happen after reconstruction.
Therefore, instead of processing events, ALICE processes time frames of currently around 2.8\,ms of continuous data.
Compared to Run 2, the Pb--Pb read-out rate increases by up to a factor of 100 at a nominal interaction rate of 50\,kHz Pb--Pb.
This makes online data compression mandatory.
ALICE has a long history of using GPUs for online reconstruction since Runs 1 and 2~\cite{bib:hlt}.
In Run 3, ALICE operates the GPU-enabled online processing farm EPN~\cite{bib:epnfrontier} (Event Processing Nodes), equipped with 2800 GPUs (AMD MI50 and MI100) and $\sim25000$ CPU cores (AMD EPYC Rome 32-core and 48-core), for online reconstruction and data compression in the ALICE O$^2$ software framework~\cite{bib:chep2023}.

In the ALICE Run 3 computing scheme, detector raw data arrive at up to 3.5 TB/s in the FLP (First Level Processor) computing farm, which performs a first level of data reduction on FPGAs.
After event building, the EPNs perform calibration, reconstruction, and high-level compression of the data, storing CTFs (Compressed Time Frames) to a temporary disk buffer at CERN and to tapes.
The detector with by far the largest data volume is the TPC.
Thus, TPC data compression~\cite{bib:ctd2019} is the most relevant and computing-intensive task online.
The full TPC processing, and therefore the majority of online processing, takes place on GPUs.
When there is no beam in the LHC, the EPN online farm is used as a GRID site for offline processing.

\section{Online Processing on GPUs}

TPC processing (clusterization, tracking, data compression) represents around 99\% of the online reconstruction workload, and more than 90\% of the total online processing workload including calibration, quality control, event building, and all orchestration and monitoring tasks.
For the TPC online processing workload, the EPN servers have around 90\% of their total compute capacity in GPUs (90\% is measured with the TPC online reconstruction workload of high-rate Pb--Pb, and may differ for other workloads with a different relative GPU speedup).
With this, the ALICE online processing of Pb--Pb data is fully GPU-bound.
The main CPU task is to feed data to the GPUs fast enough to sustain full GPU utilization.
All improvements to the GPU code translate directly to a speedup of the EPN online processing.
In the online workload, one EPN GPU replaces around 80 EPN CPU cores~\cite{bib:chep2023}.
This is measured for the full online TPC reconstruction, which is more than 90\% of the workload.
A CPU-only online processing farm would thus require more than $2800 * 80 = 224000$ CPU cores, which would be prohibitively expensive.

\section{Offline Processing on the EPN farm}

When there is no beam in the LHC, the EPN farm does not remain idle but performs offline processing.
With the majority of the compute power in GPUs, it is natural to leverage these GPU resources for offline reconstruction.
Due to the more heterogeneous workload in offline processing, TPC represents only 50\% to 60\% of the total workload in offline processing~\cite{bib:chep2023}, compared to more than 90\% in online processing.
With 90\% of the EPN compute capacity provided by GPUs but only up to 60\% of the workload running on GPUs, offline processing on the EPNs becomes CPU-bound, instead of being GPU-bound.
In this case, increasing the GPU performance will not increase the global processing throughput at all.
It only increases the GPU idle time.
While being CPU-bound, the speedup is defined by the fraction of the workload that is offloaded to the GPU.
ALICE has shown~\cite{bib:chep2023} that by offloading TPC offline reconstruction representing 60\% of the total workload to GPU, with only 40\% of the CPU workload remaining, the resulting speedup is exactly the expected factor 2.5x.
Note that all PCIe data transfers are fully pipelined, and the CPU overhead of driving the GPUs is negligible, amounting to only a fraction of a CPU core.

\begin{figure}[h]
\centering
\includegraphics[width=1.0\textwidth,clip]{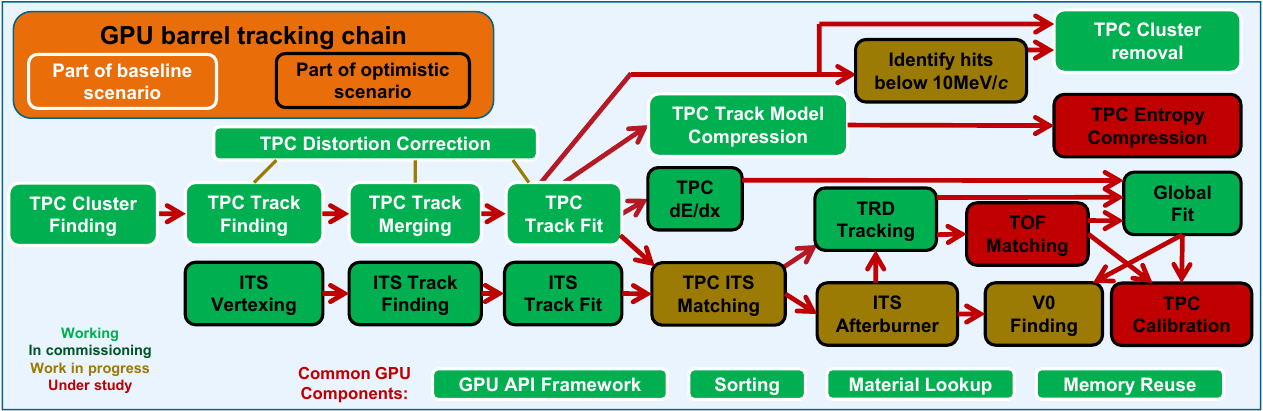}
\caption{Processing graph of the ALICE barrel tracking chain. White labels indicate tasks required in the baseline scenario running only TPC online processing on GPUs. Black labels indicate the additional tasks intended for GPU offload in the optimistic scenario to offload the full barrel tracking. Colors indicate the status of the GPU implementation of the tasks, with green indicating the tasks that can run on GPU today.}
\label{fig:barreltracking}
\end{figure}

In order to increase the GPU utilization in offline processing, ALICE follows the long-term strategy to offload the full barrel tracking to GPU (see Figure~\ref{fig:barreltracking}).
The full barrel tracking chain represents roughly 80\% of the total workload, both for pp and for Pb--Pb.
Note that this number is lower than the 90\% of the compute capacity provided by GPUs.
However, it is not guaranteed that all algorithms offloaded to GPU will see the same speedup as TPC online reconstruction.
Thus moving 80\% to the GPU is considered a realistic goal to achieve near optimal GPU utilization without becoming GPU-bound.

Nowadays, ALICE has offloaded the TPC track model decoding~\cite{bib:gabrielelhcp} for a $\sim3$\% increase, and the ITS tracking~\cite{bib:chep2024its} yielding an additional 26\% increase of the global time frame processing throughput in offline reconstruction.

\subsection{Process Multiplicities for CPU Tasks}

\begin{figure}[h]
\centering
\includegraphics[width=1.0\textwidth,clip]{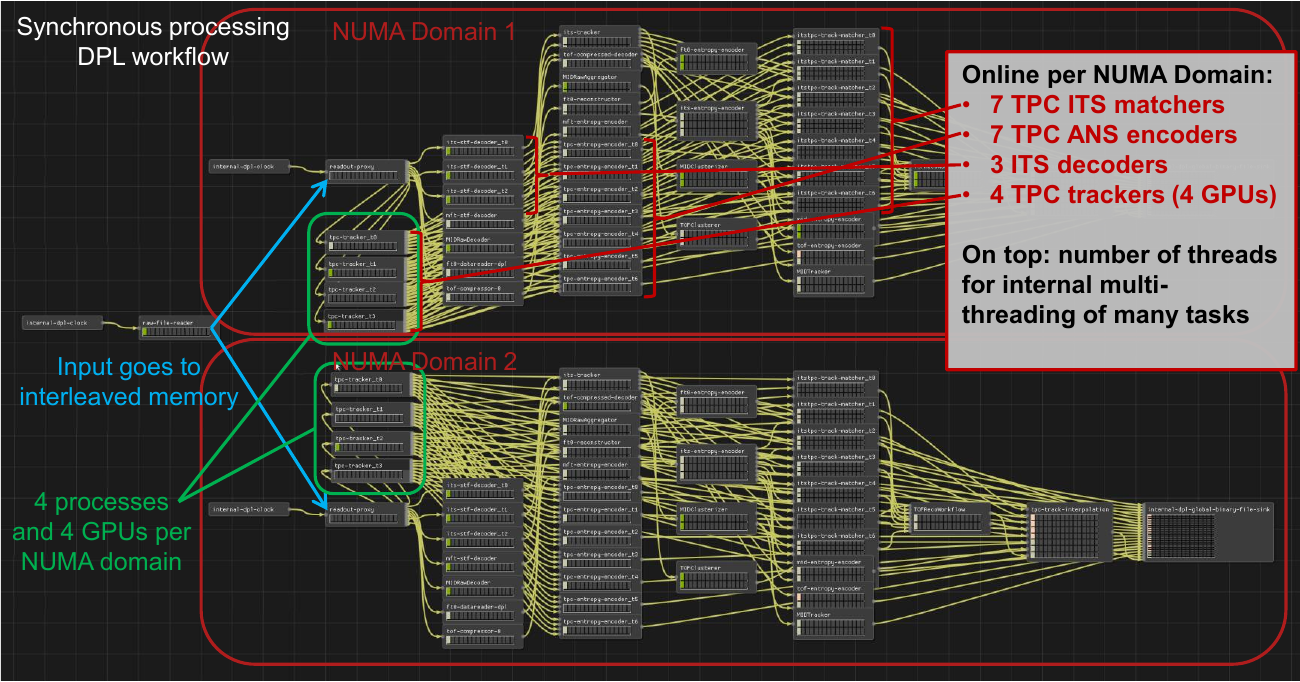}
\caption{ALICE O2 Pb-Pb Online Reconstruction Workflow Graph. Both NUMA domains run a copy of the same workflow each to avoid cross-NUMA traffic. Four tasks are running with process multiplicities, the TPC reconstruction for the four GPUs per NUMA domain and three CPU tasks.}
\label{fig:workflow}
\end{figure}

A critical aspect is to guarantee that all of the CPU processing steps are fast enough.
In the ALICE DPL (Data Processing Layer) processing framework, each task is an operating system process.
Figure~\ref{fig:workflow} shows an example.
The processing can be seen as a directed graph, in which time frames percolate through the processes.
All tasks must operate at the same global throughput.
More clearly, faster tasks are throttled to the same throughput as the slowest task.
Therefore, slower tasks run with process multiplicities: multiple instances of the same task processing the time frames in a round-robin scheme.
This avoids bottlenecks.
The total number of tasks, process multiplicities, and multi-threading inside the task must yield enough parallel processing to fully load the CPUs.
This does not mean that fast tasks are executed 100\% of the time, but rather that the slow tasks run with enough multiplicity to reach the limit of 100\% CPU load.
Unfortunately, memory constraints set a natural limit for the process multiplicities.
DPL implements rate limiting, restricting the maximum number of time frames allowed to be in flight.
This leads to a number of configurable parameters for each workflow (time frames in flight, process multiplicities), which steer the parallelism.
A fine balance has to be found, with high enough multiplicities to yield full CPU utilization, but small enough multiplicities to fit into the memory.

For online processing, one stable set of settings containing custom multiplicity values for 4 tasks is used.
Offline processing at 8-core GRID sites is also very simple, the only parameter being 5 threads for TPC reconstruction.
Offline processing with GPUs is much more complicated because it is much more heterogeneous and runs many more tasks than online.
Both pp and Pb--Pb data processing with GPUs on the EPN currently require 12 different multiplicity settings (with factors between 2 and 12).
What makes things even more complicated is the fact that the required number of instances of a task changes with software versions, data-taking conditions, and also with the set of tasks that is offloaded to GPUs.
For the future, ALICE is considering an automated tuning of the parameters based on pilot jobs.
So far, these parameters are tuned manually.
This means that all benchmark comparisons require multiple iterations of the manual tuning, to enable an apples-to-apples comparison, and to prevent one of the CPU tasks from becoming a bottleneck.

\section{Offline Processing on GPUs on the GRID}

\begin{figure}[h]
\centering
\includegraphics[width=1.0\textwidth,clip]{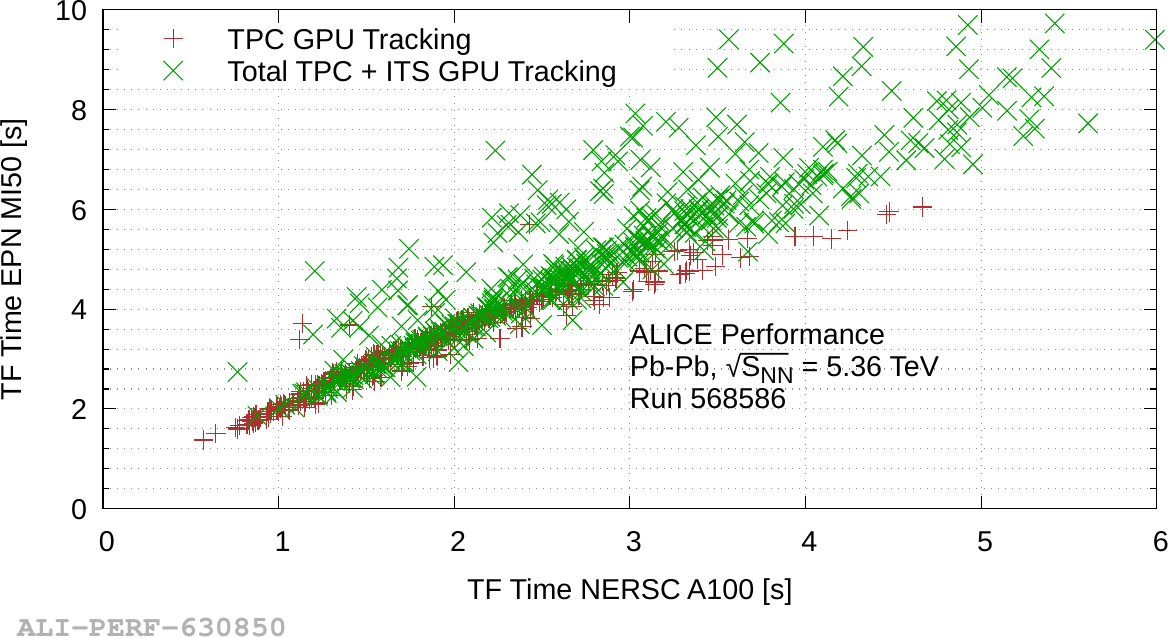}
\caption{Processing time per time frame on a single AMD MI50 on an EPN versus an NVIDIA A100 GPU at NERSC, measured in TPC-only tracking and in TPC + ITS tracking on GPU}
\label{fig:tf_time_comparison}
\end{figure}

\begin{figure}[h]
\centering
\includegraphics[width=1.0\textwidth,clip]{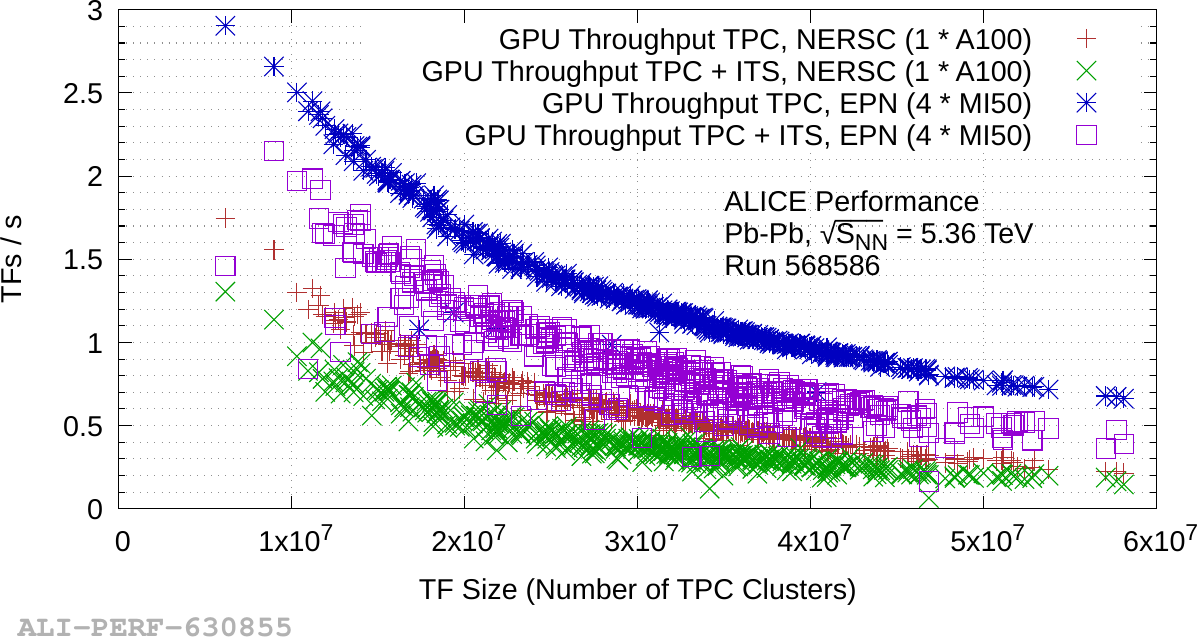}
\caption{Aggregate GPU processing throughput in time frames of jobs running on EPN (4 MI50 GPUs) and at NERSC (1 A100 GPU) performing TPC and TPC + ITS tracking on GPU versus time frame size}
\label{fig:throughput_vs_tf_size}
\end{figure}

Beyond improving the GPU usage in the EPN farm for offline processing, ALICE also aims to use GPUs at GRID sites for offline processing.
In 2026, the first successful production ran at the Perlmutter HPC cluster at NERSC equipped with NVIDIA A100 GPUs.
Figures~\ref{fig:tf_time_comparison} and~\ref{fig:throughput_vs_tf_size} compare the GPU processing speed of an EPN MI50 server with a NERSC server.
The current setup runs jobs requesting 4 MI50 GPUs and 64 virtual CPU cores on the EPN, and 1 A100 GPU and 64 virtual CPU cores at NERSC.
Figure~\ref{fig:tf_time_comparison} compares the processing time per time frame on a single GPU for TPC tracking, and for TPC + ITS tracking.
In both cases, the NVIDIA A100 GPU at NERSC is approximately twice as fast as one AMD MI50 GPU.
For large time frames, i.e. high detector occupancy and long processing times, the red curve for TPC tracking flattens, since the MI50 is more efficient at larger datasets.
The green curve for TPC + ITS tracking shows larger outliers on the MI50 GPU, since ITS processing time on the MI50 has strong fluctuations.
4 MI50 GPUs at the EPN, each having half the A100 performance, should achieve 2x the GPU throughput of 1 A100 GPU at NERSC.
Figure~\ref{fig:throughput_vs_tf_size} compares the GPU throughput of TPC and TPC + ITS tracking of jobs at NERSC and on the EPN versus the time frame size.
The figure demonstrates the expected approximately two times higher GPU throughput on the EPN.
However, since the offline processing is CPU-bound, the throughput on the GPU does not propagate to the global throughput.

\begin{figure}[h]
\centering
\includegraphics[width=1.0\textwidth,clip]{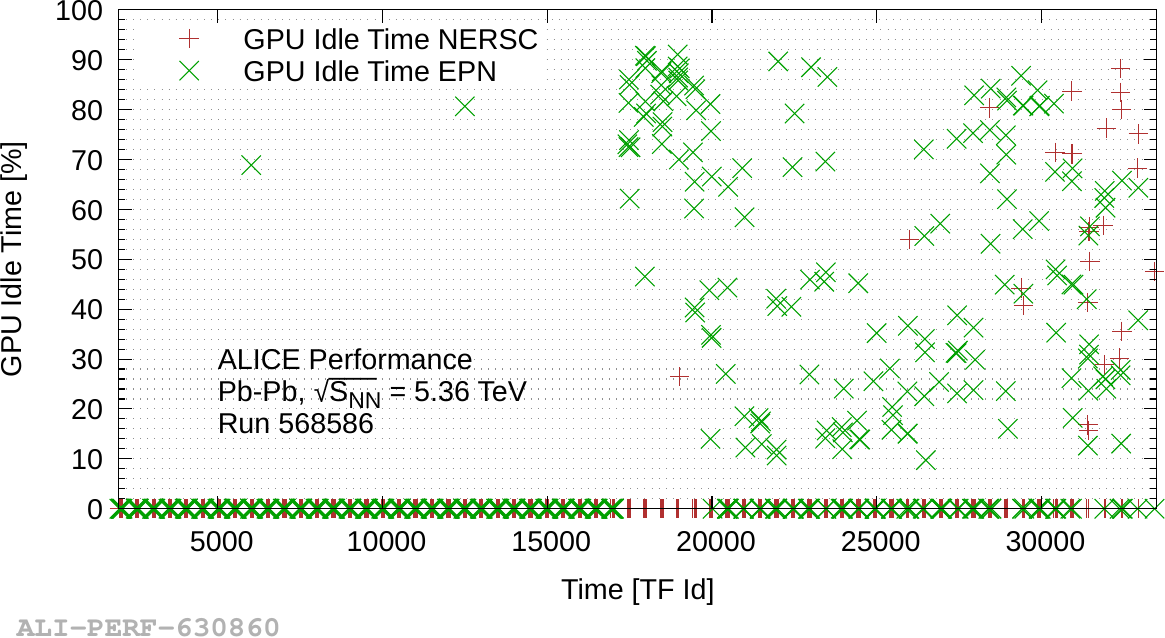}
\caption{GPU idle time of offline reconstruction jobs on EPN and at NERSC against the runtime in the job}
\label{fig:gpu_idle_time}
\end{figure}

In the CPU-bound scenario, it is important that the GPU workload is offloaded efficiently to the GPU, without creating any bottlenecks for the remaining CPU tasks.
Figure~\ref{fig:gpu_idle_time} shows the GPU idle time during the first 35000 time frames.
Note that the full production takes longer, but the behavior just continues in the same way as on the right side of the figure.
When the pipelined processing starts, the first time frames are published in an unconstrained way, until the limit for time frames in flights is reached.
The TPC and ITS tracking on GPU are located at the beginning of the processing graph, and thus they can process published time frames immediately, without waiting for the CPU.
Therefore, at the beginning the GPUs are fully loaded.
Once the DPL limit on the maximum number of time frames in flight is reached, time frame publishing gets throttled, and the GPUs begin to wait for input data.
This is visible as significant GPU idle time, with an onset once the time frame in flight limit is reached.
The job on the EPN using 4 GPUs has much faster aggregate GPU processing than 1 GPU at NERSC, thus it reaches the limit earlier, at around 17000 time frames.
At NERSC, the limit is reached after 30000 time frames.
From that time onward, the processing is fully CPU-bound on both sites with large GPU idle times, and the global throughput is fully defined by the CPU capacity.
With tuned process multiplicities, the average CPU load of the job (after some warm-up time and excluding the end when no new time frames are published and pipelines are flushed) is 90\% to 95\% (see Figure 2 of~\cite{bib:ichep2024}).

\subsection{Enabling transparent processing across GRID sites with different GPUs}

ALICE distributes its software via CVMFS to the GRID sites, and a central software build is used at all sites.
This means, in particular, that the build itself must not depend on vendor-specific GPU libraries, which may not exist at all sites.
Therefore, a plugin system is used, where all GPU backends are loaded dynamically at runtime, and only the backend plugins link to vendor libraries~\cite{bib:generic}.
This makes it possible to use one software build that can run on CPU, on NVIDIA GPUs via CUDA, on AMD GPUs via ROCm, and on compatible OpenCL devices.

Besides the different GPU backends, there are many different GPU architectures at the different sites, even for the same backend.
For optimal performance, the GPU code should be compiled for the exact architecture, while e.\;g.~CUDA also supports compiling the code to a virtual architecture.
Such code gets compiled at runtime to the actual architecture, supporting all compatible newer architectures.
ALICE aims for the highest possible performance but also for the largest possible flexibility.
The CVMFS builds are automatically created with GPU binary code for a small list of default architectures, including the 2 AMD GPU architectures used on the EPNs and a few additional ones.
For CUDA, also a broadly compatible virtual architecture target is included, which supports almost all available NVIDIA GPUs at slightly reduced performance.
OpenCL code is compiled to SPIR-V, and thus supports all compatible devices.
In order to support additional NVIDIA and AMD architectures with optimized code, ALICE O$^2$ supports RTC (Run Time Compilation) of GPU code to the local device architecture~\cite{bib:chep2024}.
The compilation process slightly increases the workflow startup time, but this is mitigated by an integrated RTC compile cache.
In this way, one central CVMFS build serves all GRID sites with all possible GPU architectures.

Another GPU-dependent aspect is a set of tuned parameters for the specific GPU architecture, like the number of threads, the number of blocks, the shared memory cache size, and a couple of algorithm-specific settings~\cite{bib:chep2024gabriele}.
The ALICE O$^2$ repository contains a list of optimized parameters, which are automatically written to parameter files that get distributed on CVMFS.
With RTC, the optimized file for the specified architecture can be read and used.

Similarly, the ALICE O2DPG (O$^2$ Data Preparation Group) scripts contain sets of optimized process multiplicities for different cases.
To simplify the deployment, these scripts also contain a mapping from the Site name to the corresponding GPU architecture and the optimized parameter file name to be used in RTC as well as the set of process multiplicities.

Finally, the JDL of the GRID job submission must specify the GPU usage and select the architecture.
In the end, this comes down to 2 settings: one integer to specify the number of GPUs for the batch job submission~\cite{bib:chep2024max}, and one string to select the correct mapping in the O2DPG scripts identified by the site name or an alias.
This means the end user has to set only 2 settings to enable the GPU processing, and decide on which GPUs to run.
Everything else is handled behind the scenes transparently.

\section{Conclusions}

In Run 3, ALICE uses GPUs heavily for online and offline processing.
Basically the full online reconstruction (~99\%) runs on GPUs.
Here, one MI50 GPU replaces $\sim80$ EPN CPU cores.
Since 2023, ALICE has been using GPUs for offline reconstruction, with a total throughput increase of up to 2.5x.
Meanwhile, TPC track model decoding and ITS tracking were ported to GPUs, for an additional increase of the total throughput by 3\% and 26\%.
Eventually, ALICE aims to have the full barrel tracking on GPUs, expecting a total throughput increase of 5x.
Online processing is GPU-bound, so the full GPU speedup is available in the application.
In contrast, offline processing is CPU-bound.
Thus, the increase of offline processing throughput cannot be computed from the GPU-speedup, but it is defined by the fraction of the workload that is offloaded from CPU to GPU.
In 2026, ALICE was the first LHC experiment to run offline reconstruction jobs on GPUs on the GRID at the Perlmutter cluster at NERSC.
The whole GPU framework is vendor-agnostic and easily configurable.
However, the need to tune CPU process-multiplicity parameters for many tasks has become a burden for supporting different GPU configurations.
ALICE is considering employing an automated tuning of these parameters in the future.
Software distribution on CVMFS and GRID job submission with GPU usage is almost fully transparent to the user.
The end user must only select the number and type of GPUs and the rest happens automatically.


\end{document}